\documentclass[twocolumn,journal]{IEEEtran}
\IEEEoverridecommandlockouts
\usepackage{tikz}
\usepackage[section]{placeins}
\usepackage{bm}
\usepackage{overpic}
\usepackage{bbm}
\usepackage{subcaption}
\usepackage{float}
\usepackage{amsmath}
\usepackage{graphicx}
\usepackage{epstopdf}
\usepackage{mathtools}
\usepackage{amssymb}
\usepackage{lettrine}
\usepackage{cite}
\usepackage{amssymb,amsmath}
\usepackage{amsgen,amsfonts,amsbsy,amsthm}
\usepackage{latexsym}
\usepackage{times}
\usepackage{algorithm}
\usepackage{algorithmic}

\allowdisplaybreaks
\makeatletter
\def\BState{\State\hskip-\ALG@thistlm}
\makeatother
\newcommand{\inprod}[2]{\left\langle #1,#2 \right\rangle}
\newcommand{\bvec}[1]{\bm{#1}}
\newcommand{\real}{\mathbb{R}}

\newcommand{\supp}[1]{\texttt{supp}\left(#1\right)}
\newcommand{\norm}[1]{\left\|#1\right\|_2}
\newcommand{\opnorm}[2]{\left\|#1\right\|_{#2}}
\newcommand{\abs}[1]{\left|#1\right|}

\newcommand{\prob}[1]{\mathbb{P}\left(#1\right)}
\newcommand{\expect}[1]{\mathbb{E}\left[#1\right]}

\newtheorem{defn}{Definition}[section]

\newtheorem{thm}{Theorem}[section]

\begin{document}
	\title{Dynamic Sample Complexity for Exact Sparse Recovery using Sequential Iterative Hard Thresholding}
	\author{Samrat Mukhopadhyay$^1$, \IEEEmembership{Member, IEEE}
		\thanks{Samrat Mukhopadhyay$^1$(samratphysics@gmail.com) is with the department of Electrical Engineering, Indian Institute of Technology, Madras.}}
	\IEEEoverridecommandlockouts
	\maketitle
	\begin{abstract}
		In this paper we consider the problem of exact recovery of a fixed sparse vector with the measurement matrices sequentially arriving along with corresponding measurements. We propose an extension of the iterative hard thresholding (IHT) algorithm, termed as sequential IHT (SIHT) which breaks the total time horizon into several phases such that IHT is executed in each of these phases using a fixed measurement matrix obtained at the beginning of that phase. We consider a stochastic setting where the measurement matrices obtained at each phase are independent samples of a sub Gaussian random matrix. We prove that if a certain dynamic sample complexity that depends on the sizes of the measurement matrices at each phase, along with their duration and the number of phases, satisfy certain lower bound, the estimation error of SIHT over a fixed time horizon decays rapidly. Interestingly, this bound reveals that the probability of decay of estimation error is hardly affected even if very small number measurements are sporadically used in different phases. This theoretical observation is also corroborated using numerical experiments demonstrating that SIHT enjoys improved probability of recovery compared to offline IHT.
	\end{abstract}	
	\section{Introduction}
	\label{sec:intro}
	\lettrine{C}{ompressed} sensing (CS) is a powerful tool in signal processing to recover unknown vectors from linear measurements using significantly smaller number of measurements than the dimension of the unknown vector. Precisely, the goal is to recover an unknown vector $\bvec{x}$ of length $N$ from a measurement vector $\bvec{y}\in \real^M$ with $M<N$ such that $\bvec{y}=\bvec{\Phi x}$. Although this is an under-determined system with infinitely many solutions, with an additional assumption that $\bvec{x}$ is $K-$\emph{sparse}, i.e., at most $K$ entries of $\bvec{x}$ are nonzero and $K<M/2$, it is possible to recover the vector $\bvec{x}$ exactly if $M=\mathcal{O}(K\ln (N/K))$~\cite{candes2006robust,candes-tao-stable-recovery,donoho2006compressed, foucart2013mathematical}. A class of popular recovery algorithms in the literature, called the \emph{greedy methods}, leverage the knowledge of $K$ to iteratively estimate the support of the unknown vector to provide fast and accurate reconstruction of $\bvec{x}$. Typically, the greedy algorithms for compressed sensing have been designed and analyzed for \emph{offline} sparse recovery where the goal is to recover a long sparse vector from a small number of given measurements along with the measurement matrix. However, an important problem to consider is \emph{online} sparse recovery, where the measurements, the measurement matrices, and even the unknown sparse vector might vary with time. A few researchers have studied this problem where they estimate time varying unknown vectors from sequential measurements~\cite{vaswani2010ls,vaswani2010modified,vaswani2016review}. These works explicitly assume that the unknown sparse vector either has a slowly varying support or the nonzero values of the signal slowly change over the same support. The associated algorithms also require a very good approximation of the initial sparse vector to start with which might not always be readily available. A less costly variant of online sparse recovery considers sparse recovery on-the-fly with sequentially arriving measurement matrices, and corresponding measurements. The work which closely addresses such problem is the adaptive filtering approach to compressed sensing~\cite{jin2010stochastic}. Here the researchers address the problem of estimating a single vector from a sequence of single measurements using a sparsity promoting regularizer, typically approximating the $l_0$ or similar non-convex sparsity promoting functions. However, apart from a heuristic study in~\cite{das2013sparse}, there does not seem to be any work which has looked at this problem systematically, when the algorithm has the explicit knowledge of the sparsity level of the unknown vector. 
	\paragraph{Contributions} In this paper, we address the problem of recovering a single unknown sparse vector with known sparsity level from sequentially obtained measurement matrices and corresponding measurements. For this, we consider extending the iterative hard thresholding (IHT), which is a very simple yet powerful iterative recovery method which estimates the support of $\bvec{x}$ at every time step to be the magnitude-wise top $K$ entries of a vector obtained from a gradient descent update~\cite{daubechies2004iterative,herrity2006sparse,blumensath2009iterative}. We analyze the probability of estimating an approximation of the unknown vector up to a given accuracy level after a time horizon of $T$ iterations. We prove that even by using very small number of measurements sporadically at different instants, one can achieve recovery guarantees similar to the one obtained for recovery with fixed measurement matrix. To the best of our knowledge, this is the first time that such sample complexity guarantees are provided for the problem of online compressed sensing, generalizing the corresponding results for the offline setting.
	%
	\section{System Model}
	\label{sec:seq-iht}
	%
	We consider $T$ slotted time instants $t=1,\cdots, T$, broken into $s$ consecutive \emph{phases} $0,1,\cdots, s-1$, associated with the time intervals $\{1,\cdots,t_1\},\{t_1+1,\cdots, t_2\},\cdots, \{t_{s-1}+1,\cdots, T\}$, respectively, where the time instants $t_0:=1\le t_1<t_2<\cdots<t_{s}:=T$ are predetermined. At time $t_i$, the measurement matrix $\bvec{\Phi}_{i+1}$ and the corresponding measurement vector $\bvec{y}_{i+1}=\bvec{\Phi}_{i+1}\bvec{x}$ for phase $i$ are received, where $\bvec{x}$ is the unknown vector of length $N$ and sparsity $K$. This sequential measurement model can be succinctly described as below:
	\begin{align}
		\bvec{y}_{t} & = \bvec{\Phi}_{t}\bvec{x}, \bvec{\Phi}_{t}\in \real^{M_{t}\times N},\nonumber\\
		\left.\begin{array}{ccc}
			\bvec{y}_t  & = & \bvec{y}_{i+1}\\
			\bvec{\Phi}_t & = & \bvec{\Phi}_{i+1}\\
			M_t & = & M_{i+1}
		\end{array}\right\}, & t_{i}+1\le t\le t_{i+1},\ 0\le i\le s-1.
	\end{align}
	The pair $\{\bvec{\Phi}_{i+1},\bvec{y}_{i+1}\}$ is used to run conventional IHT algorithm for the phase $i$. We call this setup \emph{sequential IHT }(SIHT) and describe it  in Table~\ref{tab:seq-iht}.  
	%
	%
	%
	\begin{algorithm}[t!]
		\caption{Sequential IHT}
		\label{tab:seq-iht}
		\begin{algorithmic}[1]
			\REQUIRE $0=t_0< t_1<t_2<\cdots< t_s =T$, $\bvec{x}^0$.
			\FOR {$i=0,1,\cdots, s-1$}
			\STATE Receive $\bvec{\Phi}_{i+1}$ and $\bvec{y}_{i+1}=\bvec{\Phi}_{i+1}\bvec{x}$. 
			\FOR {$t=t_i+1,\cdots t_{i+1}$}
			\STATE $\bvec{x}^{t} = H_K\left(\bvec{x}^{t-1} + \bvec{\Phi}_{i+1}^\top(\bvec{y}_{i+1}-\bvec{\Phi}_{i+1}\bvec{x}^{t-1})\right)$
			\ENDFOR
			\ENDFOR
		\end{algorithmic}
	\end{algorithm}
	Similar to the offline CS setting, a key question we would like to address in this paper is the following: \emph{How should the sample size sequence $\{M_t\}$ be chosen so that, after $T$ time instants the SIHT algorithm ensures that the estimation error is smaller than a predefined threshold?}
	
	In the next section, we find a concrete answer to this question.
	%
	%
	\section{Setting and Notations}
	In order to conduct an analysis of the SIHT algorithm, we consider in this paper a stochastic setting, where the measurement matrices are independent \emph{subgaussian} matrices. For the phase $0\le i\le s-1$, we consider the measurement matrix $\bvec{\Phi}_{i+1}=\frac{\bvec{A}_{i+1}}{\sqrt{M}_{i+1}}$, where $\bvec{A}_{i+1}$ has i.i.d. subgaussian entries with parameter $c$, variance $1$ and with isotropic rows. Also, crucially, we assume that the matrices $\{\bvec{\Phi}_i\}_{i\ge 1}$ are \emph{mutually independent}.
	
	In the rest of the paper, we use in the superscript $\top$ to denote matrix or vector transpose. Also, for any vector $\bvec{x}\in \real^N$, and any subset $S\subset [N]:=\{1,2,\cdots,N\}$, we denote $\bvec{x}_S$ to be the vector with indices restricted to the subset $S$. Similarly, $\bvec{\Phi}_S$ for a matrix $\bvec{\Phi}$ is defined to be a matrix with columns of $\bvec{\Phi}$ indexed with $S$. We denote by $\bvec{I}_K$, the $K\times K$ identity matrix. Finally, we will require the definition of the \emph{restricted isometry constant (RIC)}~\cite[pp. 133]{foucart2013mathematical}:
	\begin{defn}[Restricted Isometry Constant]
		\label{defn-ric}
		For any positive integer $K$, the restricted isometry constant (RIC) of order $K$ of a matrix $\bvec{\Phi}\in \real^{M\times N}$, denoted by $\delta_K(\bvec{\Phi})$, is defined as below:
		\begin{align}
			\delta_{K}(\bvec{\Phi}) & = \max_{S\subset [N]:\abs{S}\le K}\opnorm{\bvec{\Phi}_S^\top\bvec{\Phi}_S-\bvec{I}_K}{2\to 2},
		\end{align}
		where $\opnorm{\cdot}{2\to 2}$ is the operator norm of a matrix~\cite[pp. 344]{horn2012matrix}.
	\end{defn}
	\section{Main results}
	\label{sec:main-results}
	Consider a time instant $t$ in phase $i$, i.e., $t_i + 1\le t\le t_{i+1}$. Then, using exactly the same analysis as in~\cite{foucart2011hard}, using the definition of RIC, we obtain the following:
	\begin{align}
		\label{eq:first-upper-bound}
		\norm{\bvec{x}^{t}-\bvec{x}} & \le \sqrt{3}\delta_{3K}(\bvec{\Phi}_{i+1})\norm{\bvec{x}^{t-1}-\bvec{x}}.
	\end{align}
	Since the measurement matrices change in each phase, using the inequality~\eqref{eq:first-upper-bound} iteratively one obtains the following:
	\begin{align}
		\label{eq:second-upper-bound}
		\norm{\bvec{x}^{t_{i}}-\bvec{x}} & \le 3^{t_i/2}\prod_{j=1}^{i}\delta^{\tau_j}_{3K}(\bvec{\Phi}_{j})\norm{\bvec{x}^{0}-\bvec{x}},
	\end{align}
	where $\tau_j=t_{j}-t_{j-1},j\ge 1$. 
	%
	To study the recovery performance of SIHT, we seek to understand what is the probability with which the estimates produced by SIHT decay ``rapidly''. Before proceeding to find such a guarantee, let us first define the \emph{dynamic sample complexity} $\mathcal{M}_d(\{M_j,\tau_j\}_1^s) $ associated with $s$ phases of durations $\tau_1,\cdots, \tau_s$, such that $\sum_{j=1}^s\tau_j=T$, and corresponding measurement numbers $M_j,\ 1\le j\le s$ as below: 
	\begin{align}
		\label{eq:dynamic-sample-complexity}
		\mathcal{M}_d(\{M_j,\tau_j\}_1^s) & = \frac{g_M^2}{s\bar{p}a_M},
	\end{align}
	where $p_j=\frac{\tau_j}{T},\ 1\le j\le s,\ \bar{p}=\max_{1\le j\le s}p_j,\ a_M=\sum_{j=1}^s p_jM_j$ and, $\ g_M=\prod_{j=1}^s M_j^{p_j}$. Using this definition, we establish the following result:
	\begin{thm}
		\label{thm:prob-rapid-decay}
		Let $\epsilon\in (0,1)$, and let
		\begin{align}
			\mathcal{M}_d(\{M_j,\tau_j\}_1^s)  \ge C_1\ln (6K) + C_2K\ln(3Ne/K) + C_3\ln(1/\epsilon),
		\end{align}
		where $C_1 = C_3=96/\widetilde{c}, C_2=288/\widetilde{c}$, where $\widetilde{c}$ is a constant that depend only on the subgaussian parameter $c$. Then, with probability larger than $1-\epsilon^s$, 
		\begin{align}
			\label{eq:prob-rapid-decay}
			\norm{\bvec{x}^{T}-\bvec{x}}\le \frac{\norm{\bvec{x}^0-\bvec{x}}}{2^T}. 
		\end{align}
	\end{thm}
	Theorem~\ref{thm:prob-rapid-decay} demonstrates that the probability of decay of estimation error of SIHT can be controlled simply by controlling the associated dynamic sample complexity. There are several interesting implications of Theorem~\ref{thm:prob-rapid-decay} which we discuss below:
	
	{\bf 1)} If there is only one phase using the same fixed matrix with $M$ rows, then $s=1,g_M=a_M=M$, so that $\mathcal{M}_d = M$. Then Theorem~\ref{thm:prob-rapid-decay} asserts that exact recovery is possible with high probability if $M=\mathcal{O}(K\ln(N/K))$. This recovers the necessary sample complexity result for offline compressed sensing~\cite{candes2006robust}.
	
	{\bf 2)} If all measurement matrices used in the different phases use the same number of measurements, i.e., if $M_j=M,\ 1\le j\le s$, then, $g_M=a_M=M$ and $\mathcal{M}_d=M$. Then, by Theorem~\ref{thm:prob-rapid-decay}, if $M=s\bar{p}\mathcal{O}(K\ln(N/K))$, SIHT ensures recovery with probability $\ge 1-\epsilon^s$. Therefore, as long as $s\bar{p}=\mathcal{O}(1)$, the corresponding sample complexity necessary for SIHT for perfect recovery is of the same order as the one for offline compressed sensing.
	
	{\bf 3)} Consider the case where, for each phase $1\le j\le s$, $p_j=1/s$ and let the number of measurements $M_j$ be chosen uniformly randomly from the range $[a,b]$ for integers $1\le a\le b$, i.e., $\prob{M=m}=1/(b-a+1),\ \forall m\in [a,b]$. Then, it can be proved that (please refer to Appendix)
	\begin{align}
		\label{eq:a-nice-lower-bound-am-gm}
		\expect{\mathcal{M}_d(\{M_j,\tau_j\}_1^s)} & > \frac{2b^2}{9(a+b)}.
	\end{align}
	Therefore, the Theorem~\ref{thm:prob-rapid-decay} implies that as long as $\frac{2\alpha }{9(\alpha + 1)}b =\mathcal{O}(K\ln(N/K))$, ($\alpha=b/a$) one achieves the same convergence result as the conventional IHT with fixed measurement matrix. This is a striking result as it implies that the recovery probability is essentially unaffected, even if we are using different matrices, with different number of measurements at different phases! 
	\section{Numerical Experiments}
	\label{sec:numerical-experiments}
	%
	%
	\begin{figure}[t!]
		\centering
		\begin{subfigure}{0.33\textwidth}
			\centering
			\begin{overpic}[width=2in, height=1.5in]{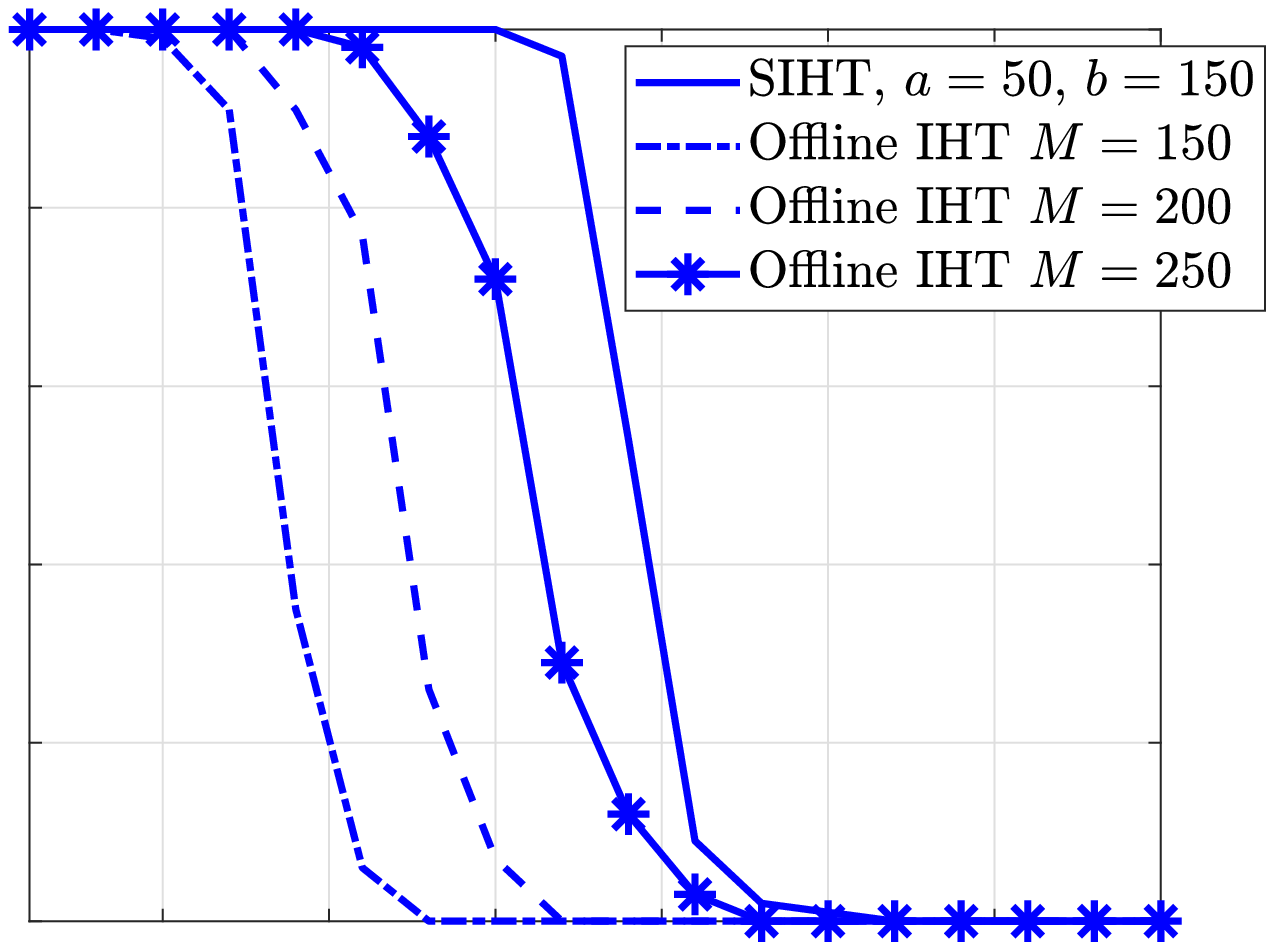}
				\put(-6,4){\rotatebox{90}{\footnotesize{\textrm{Probability of Exact Recovery}}}}
				\put(4,19){\footnotesize{0.2}}
				\put(4,31){\footnotesize{0.4}}
				\put(4,43){\footnotesize{0.6}}
				\put(4,55){\footnotesize{0.8}}
				\put(8,67){\footnotesize{1}}
				\put(21,0){\footnotesize{5}}
				\put(43,0){\footnotesize{15}}
				\put(65,0){\footnotesize{25}}
				\put(87,0){\footnotesize{35}}
				\put(50,3){\footnotesize{$K$}}
			\end{overpic}
			\subcaption{}
		\end{subfigure}
		\begin{subfigure}{0.33\textwidth}
			\centering
			\begin{overpic}[width=2in, height=1.5in]{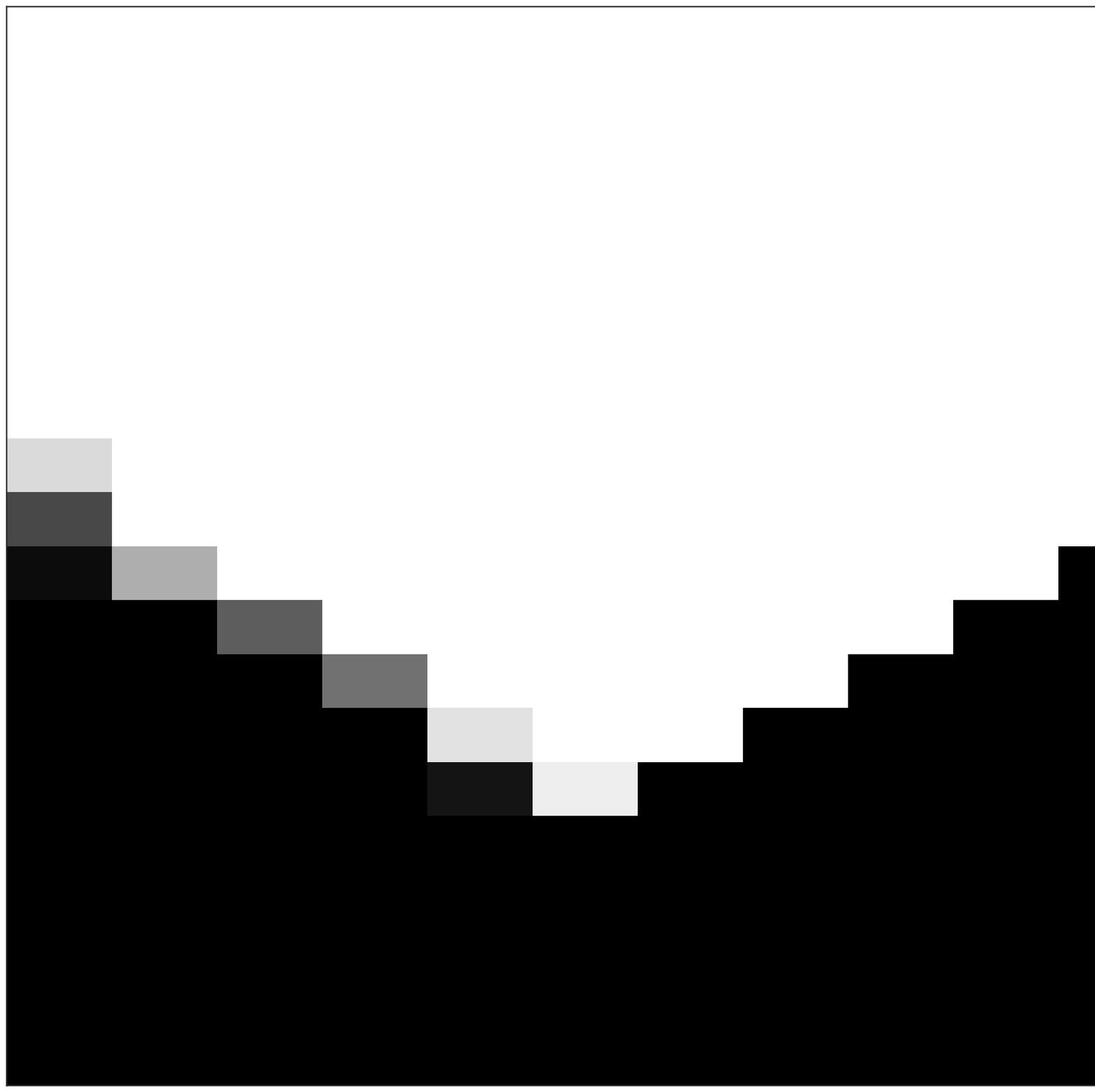}
				\put(24,-3){\footnotesize{40}}
				\put(39,-3){\footnotesize{80}}
				\put(53,-3){\footnotesize{120}}
				\put(69,-3){\footnotesize{160}}
				\put(84,-3){\footnotesize{200}}
				\put(47,2){$a$}
				\put(6,45){$b$}
				\put(-2,14){\footnotesize{40}}
				\put(-2,26){\footnotesize{80}}
				\put(-5,38){\footnotesize{120}}
				\put(-5,51){\footnotesize{160}}
				\put(-5,63){\footnotesize{200}}
				\put(16,70){\footnotesize{Probability of Exact Recovery}}
			\end{overpic}
			\subcaption{}
		\end{subfigure}
		\caption{\footnotesize{Performance of SIHT: (a) Probability of recovery comparison with Offline IHT, (b) Phase Transition Diagram with varying $a$ and $b$. White denotes probability of $1$ and black denotes $0$.}}
		\vspace{-5.5mm}
		\label{fig:siht-perfo}
	\end{figure}
	In our numerical experiments, we consider the scenario where $s=T$ and therefore, $p_j=1/T,\ 1\le j\le T$. In each phase, we choose $M_j$ uniformly randomly from the set $\{a,a+1,\cdots,b\}$. At each phase $j$, we consider measurement matrices with elements i.i.d.$\sim \mathcal{N}(0,1/M_j)$. In our first experiment, we plot the exact recovery probabilities of SIHT as well as offline IHT varying the sparsity $K$. To measure the exact recovery probability, we run each algorithm for $100$ independent random instances for each value of $K$, and at the end of each instance we count exact recovery if the estimation error satisfies $\norm{\bvec{x}^T-\bvec{x}}\le 10^{-3}$. The results are plotted in Fig.~\ref{fig:siht-perfo}(a). The plots clearly show that exact recovery probability of SIHT with $a=20$ and $b=150$ outperforms exact recovery probability of offline IHT for a range of values of measurements, namely $M=100,200,250$, clearly exhibiting the superior recovery performance of SIHT. We also plot in Fig.~\ref{fig:siht-perfo}(b) a phase transition diagram, where the probability of recovery is shown in grayscale (with white indicating probability $1$ and black $0$). In this figure, we have varied the values of $a,b$ within the range $1$ to $200$ and found the corresponding recovery probabilities of SIHT. It can be observed from the figure that for $a,b$ around $60$ and above, exact recovery always occurs whenever $b\ge a$ is chosen. Interestingly, even for $a$ below around $60$, by choosing $b$ sufficiently larger, SIHT enjoys high probability of recovery. This is in contrast to the conventional phase transition diagram in offline CS, where high probability of exact recovery is enjoyed only when the sample complexity is larger than a critical value. 
	%
	\section{Conclusions}
	\label{sec:conclusion}
	In this paper we have considered a simple extension of the conventional IHT algorithm to propose the sequential IHT algorithm for exact recovery of a sparse vector from measurement matrices and corresponding measurements arriving sequentially in phases. In Section~\ref{sec:main-results} we have introduced the notion of the  dynamic sample complexity $\mathcal{M}_d(\{M_j,\tau_j\}_1^s)$ and have proved, assuming a stochastic setting, that if $\mathcal{M}_d(\{M_j,\tau_j\}_1^s)=\mathcal{O}(K\ln(N/K))$, then SIHT can recover an approximation of the true sparse vector with arbitrary accuracy with very high probability. Further investigations of extensions of greedy algorithms like orthogonal matching pursuit (OMP), compressive sampling matching pursuit (CoSAMP), hard thresholding pursuit (HTP) etc. for the online sparse recovery problem can be considered for future research.
	\appendices
	\section{Proof of Theorem~\ref{thm:prob-rapid-decay}}
	\label{sec:proof-thm-prob-rapid-decay}
	We will denote $R=3K$ for brevity in the rest of the analysis. To begin with, recall the inequality~\eqref{eq:second-upper-bound} to obtain 
	\begin{align}
		\label{eq:first-lower-bound-rip-inequality}
		\prob{\norm{\bvec{x}^T-\bvec{x}}\le \frac{\norm{\bvec{x}^0-\bvec{x}}}{2^T}} & \ge \prob{\prod_{i=1}^s \delta^{p_j}_{R}(\bvec{\Phi}_j)\le \frac{1}{2\sqrt{3}}}.
	\end{align}
	Therefore, it is enough to find out a lower bound of the right hand side (RHS) of the above inequality. Before proceeding any further, let us denote $\Sigma_{R} = \{S\subset [N]: \abs{S}=R\}$, and $\mathcal{B}^N_{S} = \{\bvec{z}\in \real^N:\norm{\bvec{z}}\le 1,\ \supp{\bvec{z}}=S\}$.
	Recall Definition~\ref{defn-ric} of RIC to write, 
	\begin{align}
		\delta_{R}(\bvec{\Phi}) & = \max_{S\in \Sigma_R}\opnorm{\bvec{\Phi}_S^\top\bvec{\Phi}_S-\bvec{I}_K}{2\to 2}\nonumber\\
		\label{eq:expected-ric-prelim-upper-bound}
		\ & = \max_{S\in \Sigma_R}\sup_{\bvec{z}\in \mathcal{B}^N_S}\abs{\norm{\bvec{\Phi z}}^2-\norm{\bvec{z}^2}}.
	\end{align}
	Fix any subset $S\in \Sigma_R$. 
	A standard reduction argument using $\epsilon-$nets allows one to obtain (see the analysis in~\cite{foucart2013mathematical} between Eqs. (9.11) and (9.12)),
	\begin{align}
		\delta_R(\bvec{\Phi}_j) & \le 2\max_{S\in \Sigma_R}\max_{\bvec{u}\in U_S}\abs{\norm{\bvec{\Phi}_j\bvec{u}}^2- \norm{\bvec{u}}^2},
	\end{align} 
	where $U_S$ is finite set (a $1/4$-net of $\mathcal{B}_S^N$ to be precise~\cite[$\S$ 4]{vershynin2018high}) of cardinality $\abs{U_S}\le 9^R$.
	Using $\bvec{\Phi}_j=\frac{\bvec{A}_j}{\sqrt{M}_j}$, and denoting the rows of $\bvec{A}_j$ as $\{(\bvec{a}_i^{(j)})^\top\}_{i=1}^{M_j}$, one can write, 
	\begin{align}
		\norm{\bvec{\Phi}_j \bm{u}}^2 - \norm{\bvec{u}}^2
		& = \frac{1}{M_j}\sum_{i=1}^{M_j}X_{ij}(\bvec{u}),
	\end{align}
	where, for a given $\bvec{u}\in U_{S_j}$, $X_{ij}(\bvec{u})=\abs{\inprod{\bvec{a}_i^{(j)}}{\bvec{u}}}^2 - \norm{\bvec{u}}^2$. 
	Therefore, for given $p_j=\frac{\tau_j}{T},\ 1\le j\le s$, one can write, 
	\begin{align}
		\lefteqn{\prod_{j=1}^s \delta_{R}^{p_j}(\bvec{\Phi}_j) \le  \prod_{j=1}^s 2^{p_j}\max_{S_j\in \Sigma_R, \bvec{u}\in U_{S_j}}\abs{\frac{\sum_{i=1}^{M_j}X_{ij}(\bvec{u})}{M_j}}^{p_j}} & &\nonumber\\
		\ & = \frac{2}{\prod_{j=1}^s M_j^{p_j}} \max_{\substack{S_j\in \Sigma_R,\bvec{u}_j\in U_{S_j}\\1\le j\le s}}\prod_{j=1}^s\abs{W^{(j)}_{\bvec{u}_j,S_j}}^{p_j},
	\end{align}
	where $W^{(j)}_{\bvec{u},S}=\sum_{i=1}^{M_j}X_{ij}(\bvec{u})$.
	Therefore, we find that 
	\begin{align}
		\lefteqn{\prob{\prod_{j=1}^s \delta_{R}^{p_j}(\bvec{\Phi}_j) > \frac{1}{2\sqrt{3}}}} & & \nonumber\\
		\ & \le \prob{\max_{\substack{S_j\in \Sigma_R,\bvec{u}_j\in U_{S_j}\\1\le j\le s}}\prod_{j=1}^s\abs{W^{(j)}_{\bvec{u}_j,S_j}}^{p_j} >\frac{g_M}{4\sqrt{3}}}\nonumber\\
		\ & \le \sum_{\substack{S_j\in \Sigma_R, \bvec{u}_j\in U_{S_j}\\ 1\le j\le s}} \prob{\prod_{j=1}^s\abs{W^{(j)}_{\bvec{u}_j,S_j}}^{p_j} >\frac{g_M}{4\sqrt{3}}},
	\end{align}
	where in the last step we have used the union bound.
	Now, given $S_j\in \Sigma_R,\bvec{u}_j\in U_{S_j},\ 1\le j\le s$, we use the AM-GM inequality to obtain $\prod_{j=1}^s\abs{W^{(j)}_{\bvec{u}_j,S_j}}^{p_j} \le \sum_{j=1}^s p_j\abs{W^{(j)}_{\bvec{u}_j,S_j}}$, which implies that
	\begin{align}
		\label{eq:an-intermediate-inequality}
		\prob{\prod_{j=1}^s\abs{W^{(j)}_{\bvec{u}_j,S_j}}^{p_j} >\frac{g_M}{4\sqrt{3}}} & \le \prob{\sum_{j=1}^s p_j\abs{W^{(j)}_{\bvec{u}_j,S_j}} > \frac{g_M}{4\sqrt{3}}}.
	\end{align}
	Before further proceeding, observe that each $X_{ij}(\bvec{u})$ is zero mean and sub-exponential, with parameters $\beta,\kappa$, determined solely by the sub-Gaussian parameter $c$ such that $\prob{\abs{X_{ij}(\bvec{u}_j)}>t}\le \beta e^{-\kappa t},\ t>0$~\cite[pp. 191]{foucart2013mathematical}. It is straightforward to show, that for any $l\ge 0$ $\expect{\abs{X_{ij}(\bvec{u})}^l}\le \beta\kappa^{-l}\Gamma(l+1)$, so that one obtains, for any $\theta\in \real$,
	\begin{align}
		\expect{e^{\theta p_jX_{ij}(\bvec{u}_j)}} &  = 1 + \theta \expect{p_jX_{ij}(\bvec{u}_j)} + \sum_{l\ge 2}\frac{\theta^l p_j^l \expect{X_{ij}^l(\bvec{u}_j)}}{l!}\nonumber\\
		\ &  \le \exp\left(\frac{\beta\theta^2p_j^2F(\abs{\theta} p_j/\kappa)}{\kappa^2}\right),
	\end{align}
	where $F(u)=\sum_{l\ge 2}u^{l-2}$. Now, the random variable $W^{(j)}_{\bvec{u}_jS_j}=\sum_{i=1}^{M_j}X_{ij}(\bvec{u}_j)$ is the sum of $M_j$ i.i.d. zero mean sub exponential random variables, each with parameters $\beta, \kappa$. Therefore, we have, 
	\begin{align}
		\lefteqn{\expect{e^{\theta p_jW^{(j)}_{\bvec{u}_j,S_j}}} = \prod_{i=1}^{M_j}\expect{e^{\theta p_jX_{ij}(\bvec{u}_j)}}} & &\nonumber\\
		\ & \le \exp\left(\frac{\beta\theta^2p_j^2M_jF(\abs{\theta} p_j/\kappa)}{\kappa^2}\right).
	\end{align}
	Consequently, for any $\theta >0$, and with $v=\frac{g_M}{4\sqrt{3}}$, from inequality~\eqref{eq:an-intermediate-inequality} we obtain, using the moment generating functions technique~\cite[$\S$7]{foucart2013mathematical} and the mutual independence of the random variables $W^{(j)}_{\bvec{u}_j,S_j}$,
	\begin{align}
		\MoveEqLeft[5]	\prob{\sum_{j=1}^s p_j\abs{W^{(j)}_{\bvec{u}_j,S_j}} > \frac{g_M}{4\sqrt{3}}}\le e^{-\theta v} \prod_{j=1}^s\expect{e^{\theta p_j\abs{W^{(j)}_{\bvec{u}_j,S_j}}}} & & \nonumber\\
		\ & \le  e^{-\theta v} \prod_{j=1}^s \expect{e^{\theta p_jW^{(j)}_{\bvec{u}_j,S_j}}+e^{-\theta p_jW^{(j)}_{\bvec{u}_j,S_j}}}\nonumber\\
		\ & \le 2^s \exp\left(-\theta v+ \frac{\beta\theta^2\sum_{j=1}^s p_j^2M_jF(\theta p_j/\kappa)}{\kappa^2}\right).
	\end{align}
	Now, choose $0<\theta<\frac{\kappa}{\bar{p}},$, where $\bar{p}=\max_j p_j$. Then, $F(\theta p_j/\kappa) = \frac{1}{1-\theta p_j/\kappa}$. Therefore, for any $\theta\in (0,\frac{\kappa}{\bar{p}})$,
	\begin{align}
		\lefteqn{\prob{\sum_{j=1}^s p_j\abs{W^{(j)}_{\bvec{u}_j,S_j}} > \frac{g_M}{4\sqrt{3}}}} & & \nonumber\\
		\ & \le 2^s \exp\left(-\theta v+ \frac{\beta\theta^2}{\kappa^2}\sum_{j=1}^s \frac{p_j^2 M_j}{1-\theta p_j/\kappa}\right)\nonumber\\
		\ & \le 2^s \exp\left(-\theta v+ \frac{\beta\theta^2}{\kappa^2}\frac{\bar{p}a_M}{1-\theta \bar{p}/\kappa}\right),
	\end{align} 
	where we have used $p_j\le \bar{p},\ \forall j$ and denoted $a_M=\sum_{j=1}^s p_j M_j$. Now, choose $\theta = \frac{v}{\frac{2\beta\bar{p}a_M}{\kappa^2} + \frac{\bar{p}v}{\kappa}}=\frac{\kappa^2 v/\bar{p}}{2\beta a_M + \kappa v}\in (0,\kappa/\bar{p})$. Then it follows from simple algebra that 
	\begin{align}
		\MoveEqLeft[5]\prob{\sum_{j=1}^s p_j\abs{W^{(j)}_{\bvec{u}_j,S_j}} > \frac{g_M}{4\sqrt{3}}} \le 2^s \exp\left(-\frac{\kappa^2 v^2/\bar{p}}{2(2\beta a_M + \kappa v)}\right) & & \nonumber\\
		\ & \le 2^s e^{-\frac{\widetilde{c}v^2}{2\bar{p}a_M}},
	\end{align}
	where we have used $a_M\ge g_M$ to obtain $\frac{1}{2\beta a_M + \kappa v}\ge \frac{1}{a_M(2\beta + \kappa/(4\sqrt{3}))}=\widetilde{c}/a_M$. Consequently, we obtain, 
	\begin{align}
		\MoveEqLeft[12]\prob{\prod_{j=1}^s \delta_{R}^{p_j}(\bvec{\Phi}_j) > \frac{1}{2\sqrt{3}}} \le \sum_{S_j\in \Sigma_R, \bvec{u}_j\in U_{S_j},1\le j\le s} 2^s e^{-\frac{\widetilde{c}v^2}{2\bar{p}a_M}} & &\nonumber\\
		\le \left(2\binom{N}{R}9^{R}\right)^s e^{-\frac{\widetilde{c}g_M^2}{96\bar{p}a_M}} & \le \left(2R\left(9\frac{Ne}{R}\right)^{R}\right)^s e^{-\frac{\widetilde{c}g_M^2}{96\bar{p}a_M}}.
	\end{align}
	Therefore, for $\epsilon\in (0,1)$, we can ensure that $\prob{\prod_{j=1}^s \delta_{3K}^{p_j}(\bvec{\Phi}_j) \le  \frac{1}{2\sqrt{3}}}\ge 1-\epsilon^s$, if 
	\begin{align}
		\ & \left(6K\left(\frac{3Ne}{K}\right)^{3K}\right)^s e^{-\frac{\widetilde{c}g_M^2}{96\bar{p}a_M}} \le \epsilon^s.
	\end{align}
	The desired condition now follows after taking logarithms of both sides of the above.
	\section{Proof of inequality~\eqref{eq:a-nice-lower-bound-am-gm}}
	\label{sec:proof-of-a-nice-lower-bound-am-gm}
	Using AM-GM inequality, $\expect{g_M}=\expect{\prod_{j=1}^s M_j^{1/s}}=\left( \expect{M_1^{1/s}}\right)^s=\left(\expect{M_1^{1/s}}\right)^s=\left(\frac{\sum_{m=a}^bm^{1/s}}{b-a+1}\right)^s\ge \left(\prod_{m=a}^b m^{1/s}\right)^{\frac{s}{b-a+1}}=\left(\frac{b!}{(a-1)!}\right)^{\frac{1}{b-a+1}}$. Now, we use Stirling's inequalities, $n! = \sqrt{2\pi}n^{n+\frac{1}{2}}e^{-n+r(n)}$, where $\frac{1}{12n+1}\le r(n)\le \frac{1}{12n}$ to obtain, for $a\ge 2$, 
	\begin{align}
		\ & \ln\left(\frac{b!}{(a-1)!}\right) \ge (b+\frac{1}{2})\ln b - (a-\frac{1}{2})\ln(a-1)\nonumber\\
		\ & - (b-a+1) + \frac{1}{12b+1} - \frac{1}{12(a-1)}\nonumber\\
		\ & >(b-a+1)\ln b - (b-a+1) - \frac{12(b-a+1)-1}{12b\left(12(a-1)+1\right)}\nonumber\\
		\ & > (b-a+1)\ln b - (b-a+1) - \frac{b-a+1}{26}
	\end{align}
	where the penultimate inequality uses the fact that $b>a-1$ and the last inequality uses $b\ge a\ge 2$.
	Therefore,
	\begin{align}
		\frac{\ln\left(\frac{b!}{(a-1)!}\right)}{b-a+1} & > \ln b - \frac{27}{26}\nonumber\\
		\implies \expect{g_M}\ge \left(\frac{b!}{(a-1)!}\right)^{\frac{1}{b-a+1}} & > be^{-\frac{27}{26}}>\frac{b}{3}.
	\end{align}
	On the other hand, $\expect{a_M} = \frac{\sum_{j=1}^s M_j}{s}=\expect{M_1}=\frac{\sum_{m=a}^b m}{b-a+1}=\frac{b+a}{2}$. Now, using Cauchy-Scwartz's inequality, we have $\expect{\frac{g_M^2}{a_M}} \ge \frac{(\expect{g_M})^2}{\expect{a_M}}$.    
	Consequently, from the above, we obtain the desired inequality~\eqref{eq:a-nice-lower-bound-am-gm}.
	\newpage
	\bibliography{seq-iht}

\begin{thebibliography}{10}
\providecommand{\url}[1]{#1}
\csname url@samestyle\endcsname
\providecommand{\newblock}{\relax}
\providecommand{\bibinfo}[2]{#2}
\providecommand{\BIBentrySTDinterwordspacing}{\spaceskip=0pt\relax}
\providecommand{\BIBentryALTinterwordstretchfactor}{4}
\providecommand{\BIBentryALTinterwordspacing}{\spaceskip=\fontdimen2\font plus
\BIBentryALTinterwordstretchfactor\fontdimen3\font minus
  \fontdimen4\font\relax}
\providecommand{\BIBforeignlanguage}[2]{{%
\expandafter\ifx\csname l@#1\endcsname\relax
\typeout{** WARNING: IEEEtran.bst: No hyphenation pattern has been}%
\typeout{** loaded for the language `#1'. Using the pattern for}%
\typeout{** the default language instead.}%
\else
\language=\csname l@#1\endcsname
\fi
#2}}
\providecommand{\BIBdecl}{\relax}
\BIBdecl

\bibitem{candes2006robust}
E.~J. Cand{\`e}s, J.~Romberg, and T.~Tao, ``Robust uncertainty principles:
  Exact signal reconstruction from highly incomplete frequency information,''
  \emph{IEEE Trans. Inf. Theory}, vol.~52, no.~2, pp. 489--509, 2006.

\bibitem{candes-tao-stable-recovery}
\BIBentryALTinterwordspacing
E.~J. Candès, J.~K. Romberg, and T.~Tao, ``Stable signal recovery from
  incomplete and inaccurate measurements,'' \emph{Comm. Pure Appl. Math.},
  vol.~59, no.~8, pp. 1207--1223, 2006. [Online]. Available:
  \url{http://dx.doi.org/10.1002/cpa.20124}
\BIBentrySTDinterwordspacing

\bibitem{donoho2006compressed}
D.~L. Donoho, ``Compressed sensing,'' \emph{IEEE Trans. Inf. theory}, vol.~52,
  no.~4, pp. 1289--1306, 2006.

\bibitem{foucart2013mathematical}
S.~Foucart and H.~Rauhut, \emph{A mathematical introduction to compressive
  sensing}.\hskip 1em plus 0.5em minus 0.4em\relax Springer, 2013.

\bibitem{vaswani2010ls}
N.~Vaswani, ``Ls-cs-residual (ls-cs): compressive sensing on least squares
  residual,'' \emph{IEEE Trans. Signal Process.}, vol.~58, no.~8, pp.
  4108--4120, 2010.

\bibitem{vaswani2010modified}
N.~Vaswani and W.~Lu, ``Modified-cs: Modifying compressive sensing for problems
  with partially known support,'' \emph{IEEE Trans. Signal Process.}, vol.~58,
  no.~9, pp. 4595--4607, 2010.

\bibitem{vaswani2016review}
N.~Vaswani and J.~Zhan, ``Recursive recovery of sparse signal sequences from
  compressive measurements: A review,'' \emph{IEEE Trans. Signal Process.},
  vol.~64, no.~13, pp. 3523--3549, July 2016.

\bibitem{jin2010stochastic}
J.~Jin, Y.~Gu, and S.~Mei, ``A stochastic gradient approach on compressive
  sensing signal reconstruction based on adaptive filtering framework,''
  \emph{IEEE J. Sel. Topics Signal Process.}, vol.~4, no.~2, pp. 409--420,
  2010.

\bibitem{das2013sparse}
R.~L. Das and M.~Chakraborty, ``Sparse adaptive filtering by iterative hard
  thresholding,'' in \emph{2013 Asia-Pacific Signal and Information Processing
  Association Annual Summit and Conference}.\hskip 1em plus 0.5em minus
  0.4em\relax IEEE, 2013, pp. 1--6.

\bibitem{daubechies2004iterative}
I.~Daubechies, M.~Defrise, and C.~De~Mol, ``An iterative thresholding algorithm
  for linear inverse problems with a sparsity constraint,''
  \emph{Communications on Pure and Applied Mathematics: A Journal Issued by the
  Courant Institute of Mathematical Sciences}, vol.~57, no.~11, pp. 1413--1457,
  2004.

\bibitem{herrity2006sparse}
K.~K. Herrity, A.~C. Gilbert, and J.~A. Tropp, ``Sparse approximation via
  iterative thresholding,'' in \emph{2006 IEEE International Conference on
  Acoustics Speech and Signal Processing Proceedings}, vol.~3.\hskip 1em plus
  0.5em minus 0.4em\relax IEEE, 2006, pp. III--III.

\bibitem{blumensath2009iterative}
T.~Blumensath and M.~E. Davies, ``Iterative hard thresholding for compressed
  sensing,'' \emph{Appl. Comput. Harmon. Anal.}, vol.~27, no.~3, pp. 265--274,
  2009.

\bibitem{horn2012matrix}
R.~A. Horn and C.~R. Johnson, \emph{Matrix analysis}.\hskip 1em plus 0.5em
  minus 0.4em\relax Cambridge university press, 2012.

\bibitem{foucart2011hard}
S.~Foucart, ``Hard thresholding pursuit: an algorithm for compressive
  sensing,'' \emph{SIAM J. Numer. Anal.}, vol.~49, no.~6, pp. 2543--2563, 2011.

\bibitem{vershynin2018high}
R.~Vershynin, \emph{High-dimensional probability: An introduction with
  applications in data science}.\hskip 1em plus 0.5em minus 0.4em\relax
  Cambridge University Press, 2018, vol.~47.

\end{thebibliography}
\end{document}